\documentstyle[prl,aps,epsf]{revtex}

\begin{document}

\title{Spin-flux phase  
in the Kondo lattice model with classical localized spins}
\author{D.F. Agterberg
and S.Yunoki}
\address{National High Magnetic Field Laboratory,
Florida State University,\\
1800 E. Paul Dirac Dr., Tallahassee, Florida 32310
}
\maketitle
\begin{abstract}
We provide numerical evidence that a spin-flux phase exists
as a ground state of 
Kondo lattice model with classical local spins
on a square lattice.  
This state manifests itself as a double-Q magnetic order in the
classical spins with spin density at both $(0,\pi)$ and
$(\pi,0)$ and further exhibits fermionic spin currents around an elementary 
plaquette of the square lattice.  
We examine the spin-wave spectrum of this phase. We further 
discus an extension
to a face centered cubic (FCC) lattice where a spin-flux phase may also exist.
On the FCC lattice the spin-flux phase manifests itself as a triple-Q
magnetically ordered state and may exist in $\gamma$-Mn alloys.
\end{abstract}
\vglue 1 cm

The Kondo lattice model with classical local spins has emerged as one of the 
simplest models that can account for some of the physics of the manganites
\cite{expman}  
and the cuprates \cite{expcup}. For the manganites the 
ferromagnetic Kondo lattice model gives rise to the double
exchange model which has been argued to be the relevant model to explain 
the physics in these materials \cite{zen51}. 
For the cuprates the assumption of classical
local spins is clearly unrealistic, however the antiferromagnetic Kondo 
lattice  model gives rise to many 
insights into high $T_c$ materials. For example, it has been used to understand
the appearance of $d$-wave superconductivity \cite{pin93,sch95,che912} 
and it also gives rise to incommensurate magnetic and stripe \cite{buh99}
structures that have been experimentally observed \cite{che91,tra97}.
One aspect
of this model that has recently gained interest is the appearance 
of a Berry phase in the fermion wave function that arises when 
the fermion spin is strongly pinned to the
local classical spin orientation \cite{mue96}. 
This Berry phase has been argued to give
rise to a flux phase ground state on a square lattice in the manganites 
\cite{yam98}, to an anomalous hall-effect in ferromagnets \cite{ye99}, and to 
a quantized hall conductance in Kagome lattices \cite{ohg99}. 
In this paper we examine the conditions under which this Berry phase gives rise
to novel ground state structures. In particular, we give numerical
evidence that a spin-flux phase 
appears as a ground state structure of this model
on a square lattice. 
This phase is analogous to but quite different from the 
flux phases
that are usually discussed in the context of the cuprates \cite{aff88,has89}. 
The difference arises because the latter flux phases     
exhibit a finite current around each elementary plaquette of the 
square lattice, but in our case spin currents exist (for which the up and down 
electrons have opposite currents around an elementary plaquette).   
On the square lattice
the phase discussed here has a spin-flux of $\pi$ 
through each elementary plaquette. 
In this regard, it is of interest to note that a $\pi$ 
spin-flux phase has been argued to be central
in explaining 
the normal state properties of the cuprates by John and co-workers \cite{joh93}.

In this paper we will first demonstrate numerically 
that a double-Q magnetic structure 
exists as a ground state of the Kondo lattice model. We then demonstrate that
such a state is a spin-flux state with circulating spin currents and estimate
the stability region of this phase. We also determine the 
spin-wave spectrum of the double-Q magnetic phase and demonstrate
that a spin flux phase may exist for this model on a face centered cubic  
(FCC) lattice. The spin-flux phase on the FCC lattice manifests itself as a
triple-Q magnetically ordered state and has a spin flux of $\pi/2$ through
each elementary triangular plaquette that lies in the planes having Miller
indices $(1,1,1)$ (and equivalent symmetry planes).      

The model we study here is 
\begin{equation}
{\rm H=
-t{ \sum_{\langle {\bf ij} \rangle\alpha}(c^{\dagger}_{{\bf i}\alpha}
c_{{\bf j}\alpha}+h.c.)}}
-{\rm J
\sum_{{\bf i}}
{\bf{s}}_{\bf i}\cdot{\bf{S}}_{\bf i}
+J'\sum_{\langle {\bf ij} \rangle}{\bf{S}}_{\bf i} \cdot{\bf{S}}_{\bf j}}
,
\end{equation}
\noindent where ${\rm c^{\dagger}_{{\bf i}\alpha} }$ creates an electron
at site ${\bf i}=(i_x,i_y)$ with spin projection $\alpha$,
${\bf s_i}$=$\rm \sum_{\alpha\beta}
c^{\dagger}_{{\bf i}\alpha}\mbox{\boldmath{$\sigma$}}_{\alpha\beta}c_{{\bf
i}\beta}$ is the spin of the mobile electron, the  Pauli
matrices are denoted by $\mbox{\boldmath{$\sigma$}}$,
${\bf{S}_i}$ is the localized
spin at site ${\bf i}$,
${ \langle {\bf ij} \rangle }$ denotes nearest-neighbor (NN)
lattice sites,
${\rm t}$ is the NN-hopping amplitude for the electrons,
${\rm J}$ is a coupling between the spins of
the mobile and localized degrees of freedom,
and ${\rm J'>0}$ is a direct AF coupling
between the localized classical spins.
Throughout this article the unit of energy will correspond to $t=1$.
For the numerical studies a Monte Carlo technique was used.
This involves no ``sign problems'' so that by this procedure
temperatures as low as T=0.005 at
any density can be reached.
The present study has
been performed mostly on 6$\times$6
lattices with periodic boundary conditions
(PBC), but occasional runs were made
also using open and antiperiodic BC
as well as different lattice sizes (up to 12$\times$12 lattices).
The specific numerical technique used here involves a standard Metropolis
algorithm for the classical spins and an exact diagonalization for the
itinerant electrons. The details of the method are
described in Ref.~\cite{yuno}.

\begin{figure}
\epsfxsize=2.5 in
\epsfbox{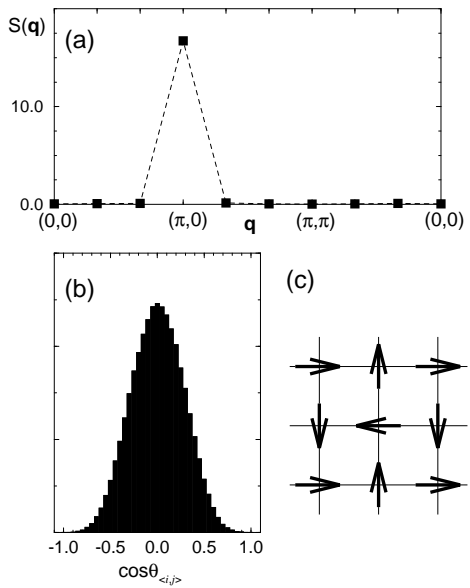}
\caption{(a) Structure factor $S({\bf q})$, (b) histogram of snapshots for the 
nearest neighbor spin
dot product, (c) and resulting classical spin structure 
for $JS=2$, $J^{\prime}S^2=0$, $T=1/200$, and $\langle n\rangle=0.5$
on a 6 by 6 lattice with periodic boundary conditions.}
\label{fig1}
\end{figure}

The spin-flux phase was identified numerically by studying the
classical spin structure factor which is the Fourier transform
of the static spin-spin correlation function $S({\bf q})=\frac{1}{N}
\sum_{{\bf n}, {\bf m}}e^{i{\bf q}\cdot({\bf n-m})}\langle{\bf S_n}\cdot{\bf S_{m}} \rangle$.
In particular, it was found that for various $JS$ and $J^{\prime}S^2$
in the vicinity of electron density 
$\langle n\rangle=0.5$ the structure factor
was peaked at ${\bf Q}=(0,\pi)$ (${\bf Q_y}$) and ${\bf Q}=(\pi,0)$ 
(${\bf Q_x}$)
(see Fig.~\ref{fig1})\cite{note}. 
To understand the possible ground states for a spin density with
this wave vector it is useful to look at a Ginzburg Landau 
free energy.   
The order parameter is
determined by the two vectors
${\bf M}_{0,\pi}$ and ${\bf M}_{\pi,0}$.
The free energy
can be simply constructed by noting that the relevant space group
representation transforms as a vector under spin rotations and 
as a scalar (that is as an $A_{1g}$ representation) 
under the little co-group $D_{2H}$ 
of the wavevector ${\bf Q}=(0,\pi)$. 
The most general dimensionless 
Ginzburg Landau free energy is 
\begin{equation}
F=-({\bf M}_{0,\pi}^2+{\bf M}_{\pi,0}^2)+
({\bf M}_{0,\pi}^2+{\bf M}_{\pi,0}^2)^2 +\beta{\bf M}_{0,\pi}^2
{\bf M}_{\pi,0}^2+\beta_2({\bf M}_{0,\pi}\cdot{\bf M}_{\pi,0})^2
\end{equation}
The minimization of this energy leads to three possible ground states:
(a) $({\bf M}_{0,\pi},{\bf M}_{\pi,0})=({\bf M},0)$, 
(b) $({\bf M}_{0,\pi},{\bf M}_{\pi,0})=({\bf M}_1,{\bf M}_2)$ with
${\bf M}_1\cdot{\bf M}_2=0$, 
(c) $({\bf M}_{0,\pi},{\bf M}_{\pi,0})=({\bf M},{\bf M})$.
The double-Q phase (b) we will argue below is the spin-flux phase
which in fact corresponds to a particular representation of  
flux phase proposed by Yamanaka {\it et al.} \cite{yam98}
(note that this phase does not lead to a peak in $S({\bf q})$ at
$(\pi/2,\pi/2)$ as suggested in Ref.~\cite{yam98}). 
The double-Q phase (c) corresponds to ordering only one half of
the local moments and is therefore not a likely ground state for this
model (note however that there exists numerical evidence
for this phase in a periodic Anderson model  
on a square lattice \cite{bon98}).  
To distinguish numerically which of these three phases 
corresponds to the phase found here  
the 
spin correlations were examined by evaluating 
${\bf S_i}{\cdot}{\bf S_j}=S^2\cos\theta_{\langle ij\rangle}$ 
(the spin dot  product of NN spins) for each pair of NN spins
and plotting the value of $\cos\theta_{\langle ij\rangle}$ in a 
histogram. The results are shown in Fig.~\ref{fig1} for $JS=2$
and $J^{\prime}S^2=0$. From this figure it is 
clear that NN spins are orthogonal 
which implies the double-Q order of phase (b) above.    

To understand the electronic properties of this double-Q magnetic phase 
we fix the  classical spins 
and find the 
fermion eigenstates (for the double-Q state this is reasonable because the
spin structure factor is very strongly peaked at $(0,\pi)$ and $(\pi,0)$ 
with little weight at other ${\bf q}$ values as can
be seen in Fig~\ref{fig1}).  
The classical spin orientation is given by 
${\bf S_i}=(S/2)[(-1)^{i_x}+(-1)^{i_y},(-1)^{i_x}-(-1)^{i_y},0]$.
Solving for the eigenstates of the resulting electronic
Hamiltonian results in four bands with dispersions 
\begin{equation}
\epsilon_{\bf k}= \pm\sqrt{(JS)^2+4(\cos^2k_x+\cos^2k_y)\pm2
\sqrt{2(JS)^2(\cos^2k_x+\cos^2k_y)+16\cos^2k_x\cos^2k_y}}
\end{equation}
where $(k_x,k_y)\in\{|k_x+k_y|\le\pi\}\cap\{|k_x-k_y|\le\pi\}$ 
are restricted to one half the original Brillouin zone.
The density of states (DOS) is linear in $|k|$ for $\langle n\rangle=0.5$  
which is characteristic of the Dirac spectrum that appears for $\pi$-flux 
phases \cite{aff88,joh93}. Also note that the dispersion relation is independent of the sign
of $J$; consequently if this flux phase is the ground state for positive
$J$ then it must also be the ground state for negative $J$.
In the limit $J=\infty$ the dispersion reduces to that found in 
Ref.~\cite{yam98}. 
To identify this phase as a spin-flux state the spin current from site 
$i$ to $j$ was determined 
\begin{equation}
{\bf j}_{i,j}=it\sum_{\alpha,\beta}\langle c^{\dagger}_{\alpha,i}
\mbox{\boldmath{$\sigma$}}_{\alpha,\beta}
c_{\beta,j}
-c^{\dagger}_{\alpha,j}\mbox{\boldmath{$\sigma$}}_{\alpha,\beta}c_{\beta,i} \rangle.
\end{equation}
It was found that only $[{\bf j}_{ij}]_z$ is non-zero and 
it is non-zero only for
NN sites. The 
resulting spin currents circulate neighboring plaquettes in 
opposite directions. 
The charge current was found to be zero. 
This spin current pattern implies a spin-flux of $\pi$ exists 
in each elementary plaquette.
Note that in the limit $J=\infty$ this result is intuitively clear;
in this limit the spin of the electron is tied to the local moment so that 
when the fermion travels around a plaquette the spin changes by $2\pi$ which
implies that the wavefunction changes sign.

\begin{figure}
\epsfxsize=2.5 in
\epsfbox{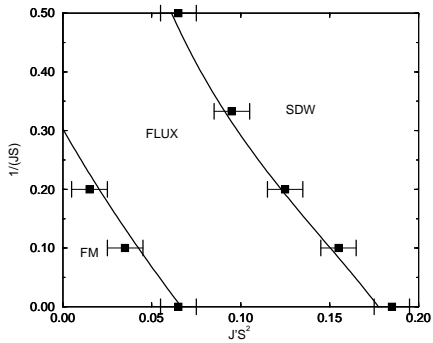}
\caption{Phase diagram for $\langle n \rangle=0.5$. The filled squares 
represent the numerical results on a $6\times6$ lattice with $T=1/200$ and with
periodic boundary conditions.}
\label{fig2}
\end{figure}

The phase diagram for $\langle n \rangle =0.5$ as a function of 
$1/(JS)$ and $J^{\prime}S^2$ is shown in Fig.~\ref{fig2}.
The solid phase boundaries were found  by comparing the energy 
of the flux phase to 
that of the canted magnetic and spin density wave (SDW) phases (the energies of 
the helical SDW phases agree with those found in Ref.~\cite{ham95}).   
At larger $J^{\prime}S^2$ the flux phase is found to be unstable to a SDW phase
characterized by ${\bf S_i}={\bf S}\sqrt{2}(-1)^{i_y}\cos(i_x\pi/2-\pi/4)$
(note this is not a helical SDW state). 
This agrees with 
the structure found numerically.  

It is of interest to determine the spin-wave spectrum
arising from the spin-flux phase. This can be done by using the 
spin-wave approximation that was introduced by Kubo and Ohata \cite{kub72}
and later used by Furukawa for the double-exchange model \cite{fur96}. 
The local spins are described by a local co-ordinate system 
in which each classical spin is
aligned along the $\hat{z}$ direction and the spin-wave operators  
$S_i^+\simeq \sqrt{2S}a_i$, $S_i^-\simeq \sqrt{2S}a_i^{\dagger}$, and
$S_i^z=S-a^{\dagger}_ia_i$ are introduced.
The spin-wave spectrum is found by keeping all $1/S$ corrections to 
the magnon self-energy.
Here we consider the limit $J=\infty$.
This results in the following effective boson Hamiltonian for the spin waves 
\begin{equation}
\sum_{{\bf k}} [\Pi({\bf k})a^{\dagger}_{{\bf k}}
a_{{\bf k}}+
A({\bf k})a^{\dagger}_{{\bf k}}a_{-{\bf k}}+h.c.]
\end{equation} 
with ${\bf k}$ summed over the whole Brillouin zone of the square lattice, 
\begin{equation}
\Pi({\bf k})=\frac{1}{2SN}\sum_{\bf q}
\left\{E_{\bf q}-\cos(\theta_{\bf q}-\theta_{{\bf k}+{\bf q}})
E_{{\bf k}+{\bf q}}
-\frac{E_{{\bf k}+{\bf q}}^2}{E_{\bf q}+E_{{\bf k}+{\bf q}}}
\left[1-\cos
(5\theta_{{\bf k}+{\bf q}}-\theta_{\bf q})\right ]\right\} 
+J^{\prime}S
(\cos k_x+\cos k_y), 
\end{equation}
\begin{equation}
A({\bf k})=\frac{1}{2SN}\sum_{\bf q}\frac{E_{{\bf k}+{\bf q}}E_{\bf q}}
{E_{\bf q}+E_{{\bf k}+{\bf q}}}
\left[\cos(2\theta_{\bf q}+2\theta_{{\bf k}+{\bf q}})
-\cos(\theta_{\bf q}-
\theta_{{\bf k}+{\bf q}})\right]+J^{\prime}S(\cos k_x+\cos k_y),
\end{equation}
$\cos\theta_{\bf k}=[\cos k_x+\cos k_y]/E_{\bf k}$,
$\sin\theta_{\bf k}=[\cos k_x-\cos k_y]/E_{\bf k}$, and $E_{\bf k}=\sqrt{2}
\sqrt{\cos^2 k_x+\cos^2 k_y}$.
The spin wave dispersion is given by 
$\omega_{{\bf k}}^2=\sqrt{\Pi({\bf k})^2-|A({\bf k})|^2}$.
The eigenstate of this mode is given by
$\delta{\bf S}_{\bf q}({\bf r})=e^{i{\bf q \cdot r}}\left[
e^{i{\bf Q_x\cdot r}}(\Pi({\bf q})+A({\bf q}))\frac{\hat{x}+\hat{y}}{2} 
+i\omega_{\bf q}\hat{z}+e^{i{\bf Q_y\cdot r}}(\Pi({\bf q})+A({\bf q}))
\frac{\hat{y}
-\hat{x}}{2}\right]$
(recall  the ordered moment is 
${\bf S}_0({\bf r})=e^{i{\bf Q_x\cdot r}}\frac{\hat{x}+\hat{y}}{2}
+e^{i{\bf Q_y\cdot r}} \frac{\hat{x}-\hat{y}}{2}$). 
The resulting spin-wave spectrum (see Fig. 3) 
agrees with the general form required by phenomenological arguments  
(found by using the method of Zhu and Walker \cite{zhu86}). 
\begin{figure}
\epsfxsize=3.5 in
\epsfbox{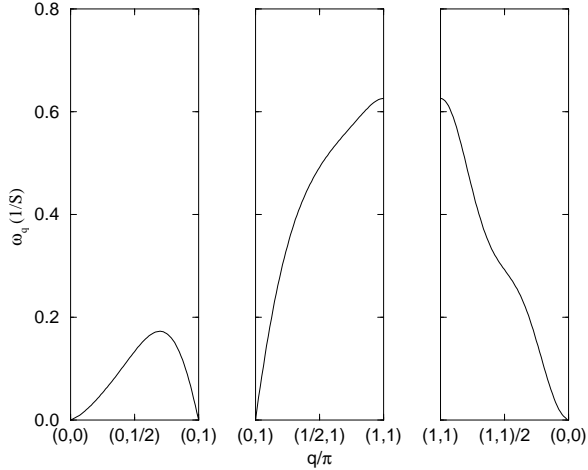}
\caption{Magnon dispersion relation for the double-Q magnetic phase
($J^{\prime}S^2=0.1$).}
\label{fig3}
\end{figure}

Given that the spin-flux phase was found to be stable on a square lattice
it is natural to ask whether such states can be realized  
on other lattice structures.  
We argue that a spin-flux phase is likely to be stable on a FCC
lattice.
For an FCC lattice two degenerate ground states of the NN 
classical Heisenberg model 
are ${\bf S}_{3Q}({\bf R})=
[(-1)^{l_1+l_2},(-1)^{l_2+l_3},(-1)^{l_2+l_3}]/\sqrt{3}$ and 
${\bf S}_{1Q}({\bf R})=[(-1)^{l_1+l_2},0,0]$ where the FCC lattice is spanned
by ${\bf R}=[l_1a(\hat{x}+\hat{y})/2,l_2a(\hat{x}+\hat{z})/2,
l_3a(\hat{y}+\hat{z})/2]$ (note that there exists a continuous degeneracy in the
ground state, but in the presence of the Kondo coupling only  
two states are relevant). 
In the limit $J=\infty$ the structure ${\bf S}_{3Q}$ gives rise to a spin-flux phase
with the spectrum 
\begin{equation}
\epsilon_{\bf k}=\pm\frac{4}{\sqrt{3}} \sqrt{\cos^2\frac{k_x}{2}\cos^2\frac{k_y}{2}+
\sin^2\frac{k_z}{2}\cos^2\frac{k_y}{2}+\sin^2\frac{k_z}{2}\cos^2
\frac{k_x}{2}}  
\end{equation}
where  the momenta are restricted to the region of the Brillouin zone 
where $-2\pi<k_z<0$. Note that for $\langle n \rangle=0.5$ the DOS
is again linear in energy.
For the structure  
${\bf S}_{1Q}$ the dispersion is  
$\epsilon_k=-4t\cos \frac{k_x}{2}\cos \frac{k_y}{2}$. Assuming that one of  
these ground states is stable (that is taking $J^{\prime}$ to be sufficiently 
large) then it is found that at $\langle n \rangle=1$ the ${\bf S}_{1Q}$ state 
is stable 
while at $\langle n \rangle=0.5$ the ${\bf S}_{3Q}$ state is stable. 
There is a transition
between these two states at $\langle n \rangle=0.7$. 
As in the case of the square
lattice the ${\bf S}_{3Q}$ phase has no net current flowing about any 
closed loops on the lattice 
but it has spin 
currents flowing around the elemental triangular plaquettes that 
exist in planes with Miller indices $(1,1,1)$ (and equivalent symmetry 
planes) . 
The spin currents that flow correspond to 
a spin-flux of $\pi/2$ per triangular elemental plaquette 
(not $\pi$ per plaquette as was the
case in the square lattice). It is intriguing to note that Hasegawa {\it et al.}
have pointed out that for a triangular lattice the optimal flux per plaquette
in a $U(1)$ flux phase is $\pi/2$ at $\langle n \rangle =0.5$ \cite{has89}.  
Both the ${\bf S}_{1Q}$ and the ${\bf S}_{3Q}$ states have been
observed in $\gamma$-Mn alloys produced by doping with Fe, Ni, or Cu 
\cite{kaw88,bia87} and it would be of interest to see if  
spin currents can be detected in the ${\bf S}_{3Q}$ phase of these materials.

In conclusion, we have given numerical evidence that a spin-flux phase
exists as a ground state of the Kondo lattice model with classical localized
spins on a square lattice. This phase gives rise to a spin-flux of $\pi$ for
electrons circulating an 
elementary plaquette of the square lattice and manifests itself as a 
double-Q magnetic order in the classical spins. We have also proposed that a
spin-flux phase may be stable on a FCC lattice. This phase manifests itself
as a triple-Q magnetic order and gives rise to a spin-flux of $\pi/2$ for
electrons circulating the elementary triangular plaquettes that lie in the 
planes with Miller indices $(1,1,1)$ (and equivalent symmetry planes).  

The authors wish to thank J.R. Schrieffer, C. Buhler, A. Moreo, 
E. Dagotto, and 
T. Hotta  for useful 
discussions.  
This work was supported by NSF DMR 9527035 and the State of Florida.


\begin{references}
\bibitem{note} This phase was called an incommensurate phase in the phase
diagram discussed previously in Ref.~\cite{yuno}.
\end{references}
\end{document}